# Exciton thermal radiation from structure-sorted carbon nanotube membranes

Akiteru Takahashi[1†], Kaichi Teranishi[1†], Shonosuke Takaichi[1], Taishi Nishihara[1,2]*, Yuhei Miyauchi[1]*

[1]*Institute of Advanced Energy, Kyoto University, Kyoto 611-0011, Japan*

[2]*Department of Physics, Tokyo University of Science, Kagurazaka 1-3, Shinjuku, Tokyo 162-8601, Japan*

*Correspondence to: nishitai@rs.tus.ac.jp (T.N.), miyauchi@iae.kyoto-u.ac.jp (Y.M.)

[†]These authors contributed equally.

## Abstract

Owing to their small binding energies, excitons in bulk semiconductors typically exhibit a sharp optical peak at low temperatures only. This limitation can be overcome by single-walled carbon nanotubes (SWCNTs) and other low-dimensional semiconductors with highly enhanced exciton binding energies. Exciton thermal radiation, which can potentially be exploited for selective thermal emission and energy harvesting, has been recently observed in individual SWCNTs heated under photoirradiation. However, whether macroscale-SWCNT assemblies can emit exciton thermal radiation under conduction heating remains unclear and constitutes an important challenge for practical applications. Herein, we observed peaked exciton thermal radiation from structure-sorted SWCNT membranes. Transmission spectroscopy showed robust exciton resonance at high temperatures, resulting in clear exciton resonance in the thermal radiation band. The absolute emissivity spectra of the membranes were determined at 850 K. Exciton dominance suppresses the contribution of thermal free carriers to the infrared absorption/emission spectra, maintaining the transparency below the optical gap even at elevated temperatures. These phenomena are not observed in bulk semiconductors, highlighting the structure-sorted SWCNT membranes as unique semiconductors leveraging stable exciton resonances at elevated temperatures.

Excitons are hydrogen-atom-like quasiparticles in which an electron and a hole are bound through an attractive Coulomb interaction. As the elementary excitations in semiconductors, excitons have long been a focal topic in photophysics research. The exciton signature is a sharp peak below the band edge in optical spectra[1]. The binding energy of excitons, the energy required for dissociating excitons into free electrons and holes, is typically of the order of a few tens of meV[1] in conventional bulk semiconductors. Consequently, distinct exciton features in the optical spectra of bulk semiconductors are difficult to observe at room temperatures (~26 meV) or higher, limiting the study and application of excitons at elevated temperatures. This historical paradigm has been shifted by the emergence of nanomaterials such as single-walled carbon nanotubes (SWCNTs; Fig. 1a)[2] and transition metal dichalcogenides[3]. Within these inherently low-dimensional materials, the carriers are spatially confined and Coulomb screening is weak, boosting the exciton binding energies to several hundred meV[4–8] and enabling the generation of thermal excitons with observable thermal radiation[9–11] (Fig. 1b). In contrast to the typical blackbody or graybody thermal radiation from bulk semiconductors, exciton thermal radiation exhibits a remarkably narrow linewidth[9,12] (Fig. 1c) originating from the atom-like well-defined energy levels of the excitons. Accordingly, the recombination of thermally excited excitons is followed by the emission of photons with an energy corresponding to the exciton energy (Fig. 1b). The exciton thermal radiation of SWCNTs in the near-infrared range has aroused particular interest because it can potentially enhance the energy efficiency of applications such as wavelength-selective emitters for thermophotovoltaic power generation[13,14]. To effectively generate electricity, a photovoltaic cell requires thermal radiation with a controlled band width from the hot emitter.

SWCNTs used as thermal light emitters in real applications must be assembled into macroscale materials and heated through conduction (left panel of Fig. 1d). To date, advances in dispersion and separation methods for each (*n*, *m*) structure of SWCNTs[15–19] have enabled the production of high-purity structure-sorted SWCNT assemblies as emerging opto-functional materials[20–30]. However, in all previous studies, the peaked light emission spectra have been measured from individual SWCNTs heated under photoirradiation or electronic current injection, which is inevitably accompanied by nonthermal excitation[9,10,31,32]. Whether purely thermal exciton radiation can be observed from a conductively heated SWCNT assemblage remains an open question. Experimentally, this has remained an open challenge due to technical difficulties in isolating purely thermal radiation from ultrathin membranes under conductive heating from thermal radiation of any other radiation sources such as heaters. Furthermore, just because exciton thermal radiation can be observed for individual SWCNTs does not necessarily mean that it can be observed in SWCNT membranes. Various factors can potentially prevent observations of exciton thermal radiation from bulk assemblies. Although individual SWCNTs exhibit sharp exciton resonance, the absorptivity contrast between the resonant and nonresonant photon energies of macroscale-SWCNT membranes can reduce with increasing membrane thickness, obscuring the exciton resonance in the thermal radiation spectra. Moreover, the exciton binding energy can be lower in assembled SWCNTs than in individual SWNCTs because the Coulomb screening in the assemblage is enhanced by the dielectric responses of the surrounding SWCNTs (right panel of Fig. 1d). At high temperatures, the exciton resonance peak is additionally broadened by enhanced exciton phonon scattering[9,10,33]. These factors can attenuate the exciton resonance in the optical spectra of assembled SWCNTs at high temperatures. Meanwhile, as the exciton resonance energy depends on chirality, SWCNT assemblies should comprise same-structure (same-chirality) SWCNTs[34,35]. A mixed-chirality SWCNT assembly exhibits a relatively featureless optical spectrum, including thermal radiation[36–42]. Therefore, determining whether exciton thermal radiation can be observed from the assemblies of structure-sorted SWCNTs under conductive heating and elucidating the high-temperature optical properties of these materials are essential for fundamental science and technological applications.

Here, we report the observations of exciton thermal radiation from structure-sorted SWCNT membranes. To this end, we prepared a dispersion enriched with SWCNTs with chiral indices of (7,5)[15] and fabricated a membrane from the dispersion via vacuum filtration methods[29]. First, we transferred the SWCNT membrane onto a sapphire substrate, heated it with a ceramic heater, and obtained its transmittance spectra. After the initial heating and cooling process, the exciton peak showed a reversible spectral change in the temperature range 300–800 K. As the temperature increased, the exciton peak exhibited red-shifting and linewidth broadening while the integrated oscillator strength remained constant, indicating high-temperature robustness of the exciton state in the SWCNT membrane. The peaked thermal radiation spectra were observed while heating the free-standing membrane using a tungsten heater, especially designed for this study. At 850 K, the spectral shapes of the emissivity peaks agreed with the predicted absorptivity, implying that the peak of the thermal radiation was contributed by excitons. The robustness of the excitons, even at 850 K, suppressed the infrared absorptivity and emissivity arising from free carriers. This situation contrasts with that in bulk semiconductors[43,44]. Our ultrathin structure-sorted SWCNT membranes present as a unique material showing the optical characteristics of nearly ideal two-dimensional sheets even at very high temperatures; meanwhile, the one-dimensional nature of the electronic states in each individual SWCNT ensures the dominance of exciton resonance. These materials are highly promising for thermal radiation control and other high-temperature photonics that can exploit their exciton resonances.

**Characterization of the SWCNT membranes**

The single-chirality SWCNT membranes were fabricated and characterized as follows (see Methods section for details). Briefly, a blue-colored dispersion enriched with (7,5) SWCNTs (Fig. 2a) was separated from the mixed samples using the organic polymer poly(9,9-dioctylfluorenyl-2,7-diyl) (PFO)[15] as a dispersant. The dispersion was vacuum filtered to fabricate the membrane, which was then transferred onto a sapphire substrate (hereafter referred to as an *on-sapphire membrane*) for optical characterizations and thickness measurements (Fig. 2b). The absorption spectrum of the dispersion (Fig. 2c) displays the clear peaks of the first ($S_{11}$), second ($S_{22}$) and third ($S_{33}$) subband exciton states of the (7,5) SWCNTs at 1.186, 1.898 and 3.614 eV, respectively[34]. The minor peak at 1.092 eV arises from the $S_{11}$ exciton state of (7,6) SWCNT[34], which are minorly contained in the dispersion. Following Ref. 45, the concentrations of (7,5) and (7,6) SWCNTs were determined as 0.66 and 0.02 µg mL$^{-1}$, respectively, clarifying that the dispersion was composed of almost pure (7,5) SWCNTs and the dispersant polymer molecules. The broad peak at 3.220 eV is attributed to absorption of the dispersant polymer. For reference, the absorption spectrum of the dispersant polymer (0.24 µg mL$^{-1}$) is also displayed in Fig. 2c. As their absorption peaks did not decrease after repeated procedures to remove the free polymers (see Methods), the residual polymers were assigned to those wrapped around SWCNTs.

Figure 2d shows the reflectance and transmittance spectra of the on-sapphire SWCNT membranes. The reflectance spectra were obtained in two configurations, one with the incident light entering from either side of the SWCNT membrane (top panel) and the other with light entering from the sapphire substrate (middle panel). The reflection spectra are distinct because the refractive index at the air–SWCNT interface differs from that at the sapphire–SWCNT membrane interface. All optical spectra display the signatures of the $S_{11}$ and $S_{22}$ exciton states of (7,5) SWCNTs at photon energies slightly below (a few meV) those in the spectrum of the dispersion. To determine the complex dielectric function (Fig. 2e) and membrane thickness (53 nm), we fitted all spectra to the model functions of the individual features in the optical spectra of SWCNTs[29] (see Supplementary Note 1 for details). The dispersion spectrum dominantly featured the $S_{11}$ exciton resonance of (7,5) SWCNTs; the peak structures of other chirality SWCNTs were absent except for a small peak attributed to residual (7,6)

SWCNTs at a photon energy below that of the $S_{11}$ exciton states. This observation further supports the high purity of the (7,5) SWCNTs.

**Temperature dependence of the transmittance spectra**

To determine the fundamental optical properties of the (7,5) SWCNT membrane at high temperatures, we examined the high-temperature exciton response via transmittance spectroscopy. For this purpose, the SWCNT membrane was placed on one side of the sapphire substrate and the opposite side of the substrate was glued to the ceramic heater (Fig. 3a). The sample temperatures were measured by a thermocouple positioned in the immediate vicinity of the SWCNT membrane. The SWCNT membrane was irradiated with collimated white light from a halogen lamp and its transmittance was measured between 300 K and 800 K throughout three cycles of heating and cooling. The transmittance spectrum was irreversibly altered during the initial heating–cooling cycle but reversibly altered during the second and third cycles (Supplementary Fig. 1). Figure 3b shows the complex dielectric functions of the SWCNT membrane at 300 K before (gray curves; also plotted in Fig. 2e) and after the heating–cooling processes (orange curves) (here, the complex dielectric functions were derived from the optical spectra shown in Supplementary Fig. 2). The exciton peaks show redshifts and broadened linewidths, and the absorption peaks of the polymers (open triangles) are diminished. Moreover, the membrane thickness reduced from 53 to 38 nm. These results indicate decomposition of the polymers and a consequent reduction in thickness. The redshift of the $S_{11}$ exciton peak indicates that as the polymer molecules decomposed, close contact among the nanotubes altered the surrounding dielectric environment.

Figure 3c shows the transmittance spectra of the on-sapphire SWCNT membrane during the third temperature cycle at different temperatures. As the temperature increased, the exciton peaks became more redshifted, their intensities reduced, and their linewidths broadened. From the transmittance spectrum at different temperatures, we obtained the complex dielectric functions ($\tilde{\varepsilon} = \varepsilon_1 + i\varepsilon_2$) at each temperature. More specifically, the transmittance spectra were fitted to a model complex dielectric function obtained by incrementally modifying the parameters describing the $S_{11}$ exciton and its phonon sideband, which were initially set at room temperature. The simulation results (Fig. 3c, inset) well reproduce the experimental results (Supplementary Fig. 3 for details). Figure 3d shows the temperature dependence of the imaginary part of the dielectric function ($\varepsilon_2$) and its corresponding real part ($\varepsilon_1$; inset). Between 400 K and 800 K, the dominant exciton peak exhibits an approximately 60% decrease in intensity and an approximately twofold increase in full-width-at-half-maximum. Linewidth broadening weakened the photon-energy dependence of the real part (inset). Figure 3e plots the temperature dependence of the optical susceptibility ($\tilde{\chi} = \chi_1 + i\chi_2$) of the $S_{11}$ exciton of the (7,5) SWCNTs contributing to the imaginary part of the complex dielectric function. The spectral shape is well reproduced by a susceptibility function based on the Lorentz model, given by $\tilde{\chi}(\omega) = f[(\omega_0^2 - \omega^2) - i\omega\gamma]^{-1}$, where $f$, $\omega_0$, and $\gamma$ are the strength, exciton resonant optical frequency, and damping term, respectively. Figure 3f–h summarizes the $\hbar\omega_0$, $\hbar\gamma$, and area of the imaginary part $\chi_2$ obtained from fitting, respectively (where $\hbar$ is the reduced Planck constant). The $\hbar\omega_0$ and $\hbar\gamma$ parameters are temperature-dependent (as observed above) but the area, which corresponds to the oscillator strength, is almost independent of temperature (Fig. 3h). Therefore, even under the aggregation conditions of the membrane, the excitons in the SWCNTs are thermally stable against dissociation into free carriers at 800 K and retain their oscillator strength.

**Observation of exciton thermal radiation**

Because exciton resonance in the complex dielectric function is retained at high temperatures, the SWCNT membrane is expected to emit exciton thermal radiation. For this observation, we prepared a free-standing SWCNT membrane on a specially designed heater, namely, a thin (0.1 mm) tungsten plate (see Fig. 4a). Figure 4b is a schematic of the optical setup for observing radiation from the SWCNT membrane. When a voltage is applied across the tungsten plate, Joule heat is mainly generated around the honeycomb-structured part. The resulting thermal energy is then conducted to the SWCNT membrane. The thermal radiation of the SWCNT membrane was detected through an optical fiber placed near the membrane to avoid detection of radiation from any other thermal radiation sources. The temperature of the SWCNT membrane was determined from the spectral shape of the emissivity spectrum obtained from the thermal radiation spectrum (see Methods for details). Briefly, using the complex dielectric functions obtained from the experiments above (Fig. 3d), we calculated the emissivity spectra in the 350–800 K range (Supplementary Fig. 4). The resulting emissivity spectra revealed that the intensity ratio of the exciton peak and its phonon side peaks is almost constant. The temperature at which the emissivity spectrum obtained from the measured thermal radiation spectrum divided by the Planck function satisfied the intensity-ratio characteristic was determined as the temperature of the SWCNT membrane. Figure 4c and d show the thermal radiation spectrum and the absolute spectral emissivity of the free-standing SWCNT membrane, respectively, at ~850 K. The thermal radiation spectrum exhibits a peak at 1.0 eV and a shoulder peak at around 1.1–1.2 eV. The emissivity spectrum similarly shows a major peak structure at 1.1–1.2 eV arising from the $S_{11}$ exciton states of the (7,5) SWCNTs, together with a small peak around 1.0 eV produced by residual (7,6) SWCNTs. From these results, the thermal radiation peaks at 1.1–1.2 eV and 1.0 eV in the thermal radiation spectrum (Fig. 4c) were attributed to exciton resonances of the (7,5) and (7,6) SWCNTs, respectively. Even under conduction heating, the excitons in the SWCNT membrane were thermally excited and recombined, and thereafter emitted photons at an energy corresponding to the exciton energy. At ~850 K, the intensity of the thermal radiation peak assigned to the $S_{11}$ exciton resonance from (7,6) SWCNTs exceeded that from (7,5) SWCNTs (Fig. 4c). This is because that (7,6) SWCNTs, having a lower exciton energy, are more easily thermally excited than (7,5) SWCNTs.

The emissivity was strongly suppressed at photon energies below the $S_{11}$ exciton states of (7,5) and (7,6) SWCNTs, even at ~850 K (Fig. 4d). In contrast, the emissivity of semiconductors is typically increased below the optical gap because free carriers are absorbed and emitted at high temperatures (e.g., the optical gap of silicon totally disappears at ~800 K[43,44]). Consequently, the radiation intensities at photon energies below the exciton states are largely suppressed from that of blackbody radiation at the same temperature (Fig. 4c, inset). Figure 5a shows the absolute spectral emissivity of the SWCNT membrane at ~850 K extrapolated to the infrared region (see Methods for the details), and Fig. 5b compares the thermal radiation spectra of the SWCNT membrane and a blackbody at 850 K on a log–linear scale. The membrane radiation is much less intense than the blackbody radiation at the lower-energy side, especially at the blackbody radiation peak at 200 meV, indicating the higher spectral selectivity of near-infrared thermal radiation from the SWCNT membrane. The proportion of integrated radiation energy above 0.67 eV (the estimated bandgap energy of the GaSb photovoltaic cell typically used in thermophotovoltaic energy conversion) which is indicated as a dotted line in Fig. 5b, is approximately 20% of the total radiated energy in the 0.05–1.36 eV range of photon energies. This spectral selectivity[46] is an order of magnitude higher than blackbody's spectral selectivity (~2%) at the same temperature. This remarkably high value as an intrinsic material is obtained without controlling the radiation field through photonic fine structures, highlighting the attractiveness of structure-sorted SWCNT membranes as wavelength-selective emitter materials in the near-infrared region. Additionally, as shown in Fig. 4d, the emissivity of the exciton resonance peak is 0.5, which coincides with the maximum emissivity and absorptivity of an ultimately two-dimensional sheet with a very high absorption coefficient[47]. This finding suggests that whereas a one-dimensional electronic system of SWCNTs ensures the thermal stability of excitons, thin SWCNT membranes behave as nearly ideal two-dimensional sheets with exceptionally strong absorption of near-infrared light. Therefore, thin SWCNT-membrane is a promising optofunctional material potentially enabling unconventional photonics strategy for controlling near-infrared thermal radiation.

In conclusion, our results highlight fundamental distinctions between the high-temperature optical properties of structure-sorted SWCNT membranes and conventional semiconductors. Owing to their unique characteristics, SWCNT membranes can sustain sharp exciton resonance even at very high temperatures, opening a potential pathway toward next-generation energy-harvesting and thermal photonics technologies.

**Methods**

**Dispersion preparation:** (7,5) SWCNTs were selectively separated from CoMoCAT SWCNTs (Sigma-Aldrich) in toluene using poly(9,9-dioctylfluorenyl-2,7-diyl) (PFO) polymer as the dispersant[15]. The PFO (40 mg) was dissolved in 40 mL toluene and then well mixed with 20 mg of the SWCNT raw material through bath sonication (60 min), followed by tip sonication (3 h) at 24 W. The mixed dispersion was ultracentrifuged at 19,000$g$ for 30 min, and the (7,5) SWCNT dispersion was obtained as the supernatant. To remove excessive free PFO molecules, the dispersion was vacuum-filtered on a hydrophilic MCE membrane filter with a pore size of 0.05 µm (JVWP02500, MERCK). The SWCNTs on the membrane filter were then re-dispersed in toluene, and their absorbance spectrum were measured in an ultraviolet–visible–near-infrared (UV–Vis–NIR) spectrometer (V-770, JASCO). For UV–Vis–NIR measurements, the sample was dispensed into an optical cell with a path length of 10 mm.

**Membrane preparation:** The SWCNT membranes were fabricated via vacuum filtration[29]. The dispersion was filtered under a vacuum pressure of 70–80 kPa on a hydrophilic mixed cellulose ester membrane filter with a pore size of 0.05 µm (VSWP02500, MERCK) and then dried at 3 kPa for 1 h. Subsequently, the pressure was elevated to 30 kPa, and toluene was added for washing. The membrane was again dried at 3 kPa for 1 h. The SWCNT membrane on the filter was cut and transferred to the sapphire, metal washer, and a tungsten plate via wet-transfer processes. The optical spectra of the on-sapphire membranes were measured in the abovementioned UV–Vis–NIR spectrometer.

**Temperature-variable transmittance spectroscopy:** The SWCNT membrane was transferred to one side of the sapphire substrate. The opposite side of the substrate was affixed to the ceramic heater using ceramic adhesive (Fig. 3a). The sample temperature was measured using a k-type thermocouple positioned in the immediate vicinity of the substrate. The SWCNT membrane was irradiated with collimated light from a halogen lamp. The transmitted light was focused onto the fibers and then measured using a spectrometer equipped with an indium gallium arsenide (InGaAs) array detector (NIRQuest+, Ocean Optics).

**Thermal radiation measurement:** The membrane was transferred to the hollow honeycomb structure of the tungsten plate, forming a free-standing SWCNT membrane (Fig. 4a). This sample was thermally treated to decompose the polymers. First, the polymers wrapped around the SWCNTs were decomposed through a rapid annealing-and-cooling procedure[48]. During this procedure, the sample was placed in a quartz tube, pressurized at 5–10 Pa, and rapidly heated to 600°C in a furnace. The sample was maintained at 600°C for 5 min and then cooled to room temperature over a 10-min period under an argon flow. Thereafter, the sample was placed in a vacuum chamber and an electric current was applied to the tungsten plate. The sample was heated by the Joule heat generated within the hollow honeycomb structure of the plate. The tungsten plate was cooled using copper blocks connected to liquid nitrogen. The thermal radiation of the SWCNT membrane was detected through a 600-μm diameter optical fiber placed in proximity to the SWCNT membrane (Fig. 4b) and then measured by the NIRQuest+ spectrometer equipped with the InGaAs array detector. The numerical aperture was increased (as much as possible) to prevent the entry of thermal radiation from the hotter tungsten plate into the fiber.

**Temperature estimation from the emissivity spectra:** The temperature of the thermally emitting SWCNT membrane (Fig. 4) was determined from the spectral shape of the emissivity. We initially examined the absorbance spectra of the SWCNT membrane, which are consistent with the emissivity spectra according to Kirchhoff's law[49]. Supplementary Figure 4a shows the absorptance spectra of the free-standing SWCNT membrane at varying temperatures. These spectra were calculated using the optical transfer matrix technique[50], where the complex dielectric function was obtained from the on-sapphire membrane (Fig. 3d). Although the exciton peak was redshifted with temperature, the shapes of the absorptance spectra were less influenced by temperature than the complex dielectric function (Fig. 3d). After normalizing the absorbance spectra at different temperatures to the exciton peak intensity (Supplementary Fig. 4b), the normalized intensity of the exciton phonon sideband around 1.3–1.4 eV was ~0.6 at all temperatures. Such spectral shape features of the emissivity spectra are expected from Kirchhoff's law. Candidate emissivity spectra at various temperatures were calculated from the thermal radiation spectra shown in Fig. 4c. Supplementary Figure 4c shows the resulting emissivity spectra with their intensities normalized at the $S_{11}$ exciton resonance (1.1–1.2 eV). The normalized absorptance of the phonon sideband is largely temperature dependent and is ~0.6 when taking the emissivity at an assumed temperature of 850 K. Supplementary Figure 4d compares the emissivity spectra at an assumed temperature of 850 K (Supplementary Fig. 4c) with the absorptances at 800 K (Supplementary Fig. 4b). To account for the 50 K temperature difference, the absorptance spectra were shifted by 5 meV to lower photon energies. The 5-meV shift was determined by linearly extrapolating the temperature dependence of the absorptance peak of the $S_{11}$ exciton (Supplementary Fig. 4d, inset) obtained from Supplementary Fig. 4a. Given the strong agreement between the two spectra, the temperature of the thermally emitting free-standing SWCNT membrane (Fig. 4c) was estimated as ~850 K.

**Calculation of spectral selectivity:** The emissivity spectrum of the SWCNT membrane was extrapolated to the lower photon-energy range using the tail of the Lorentz function to decrease the emissivity toward lower energies because the exciton oscillator intensity is preserved (Fig. 3h) and the excitons do not dissociate into free carriers.

**Acknowledgments:** This work was supported by JST CREST Grant Number JPMJCR185(Y.M.), and JST FOREST Grant Number JPMJFR222N(T.N.), and JSPS KAKENHI Grant Numbers JP22K18287(Y.M.), JP24H00044(Y.M.), and JP19K15384(T.N.), JP21K14486(T.N.), JP23H01791(T.N.).

**Author Contribution:** T.N. and Y.M. conceived the concept. A.T. prepared the samples. K.T. and A.T. arranged the optical setups for high temperature measurements, and A.T. and S.T. carried out the measurements. A.T. and T.N. analyzed the experimental data. All the authors contributed to writing the manuscript.

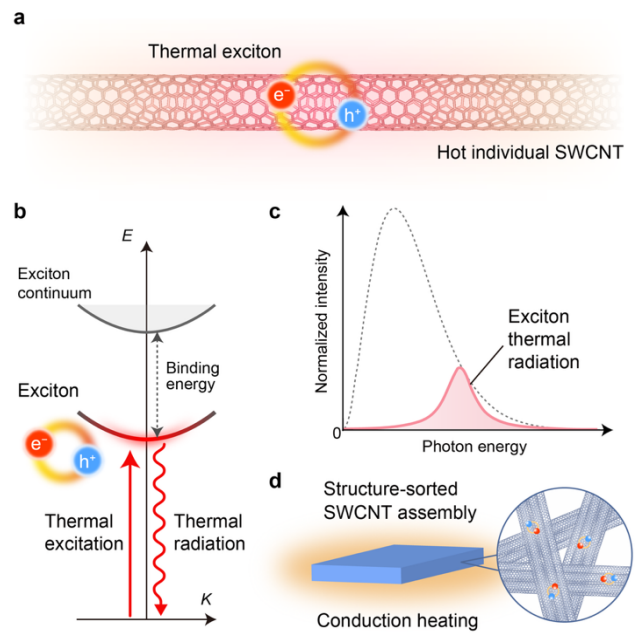

**Fig. 1 | Schematics of thermal exciton in carbon nanotubes. a** A thermal exciton (a bound electron–hole pair) in a hot individual single-walled carbon nanotube (SWCNT). **b** Exciton and continuum energy levels in SWCNTs ($E$, energy; $K$, wavenumber vector). **c** Spectral features of exciton thermal radiation from an individual SWCNT (red curve) and blackbody radiation (dotted curve). The peak intensity of exciton thermal radiation is normalized to blackbody radiation curve. **d** Excitons in a membrane of assembled structure-sorted SWCNTs.

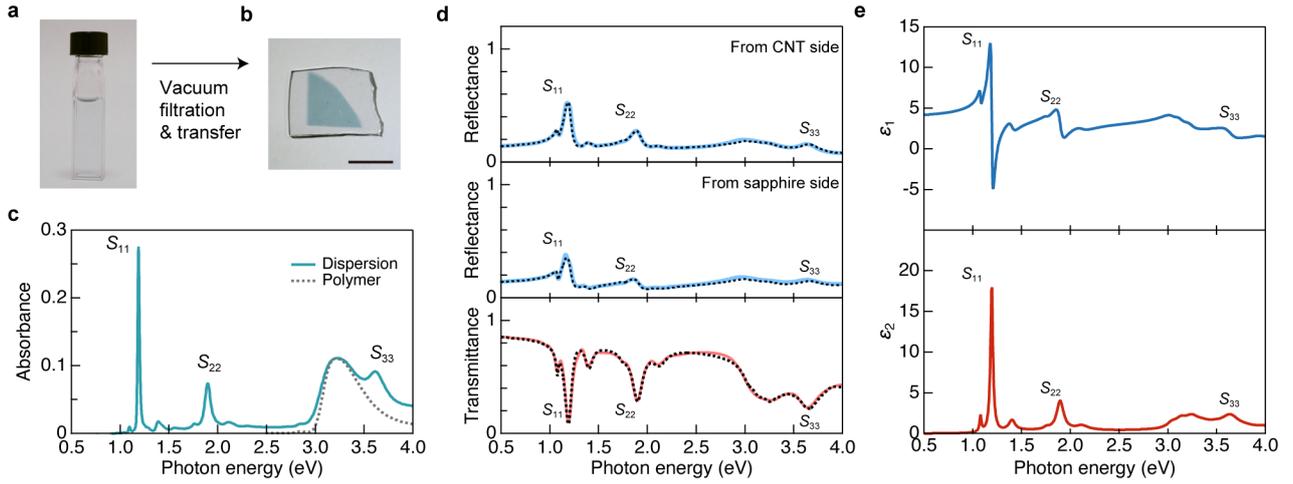

**Fig. 2 | Characterization of the structure-sorted carbon nanotube samples.** The (7,5) carbon nanotube (CNT) dispersion (**a**) and the membrane transferred onto a sapphire substrate (**b**) (scale bar = 5 mm). **c** Absorption spectrum of the (7,5) CNT dispersion in an optical cell with a path length of 10 mm. For reference, the absorption spectrum of the dispersant polymer at 0.24 μg mL$^{-1}$ (gray dotted curve) is also plotted. $S_{ii}$ is the *i*th subband of the exciton state of (7,5) CNTs. **d** Reflectance and transmittance spectra of the (7,5) CNT membrane. The reflectance spectra (blue curves) were obtained in two configurations: incident light entering from either side of the CNT membrane (top panel) and incident light entering from the sapphire substrate (middle panel). The dotted black curves are the fitting results. **e** Complex dielectric function ($\tilde{\varepsilon} = \varepsilon_1 + i\varepsilon_2$) of the (7,5) CNT membrane.

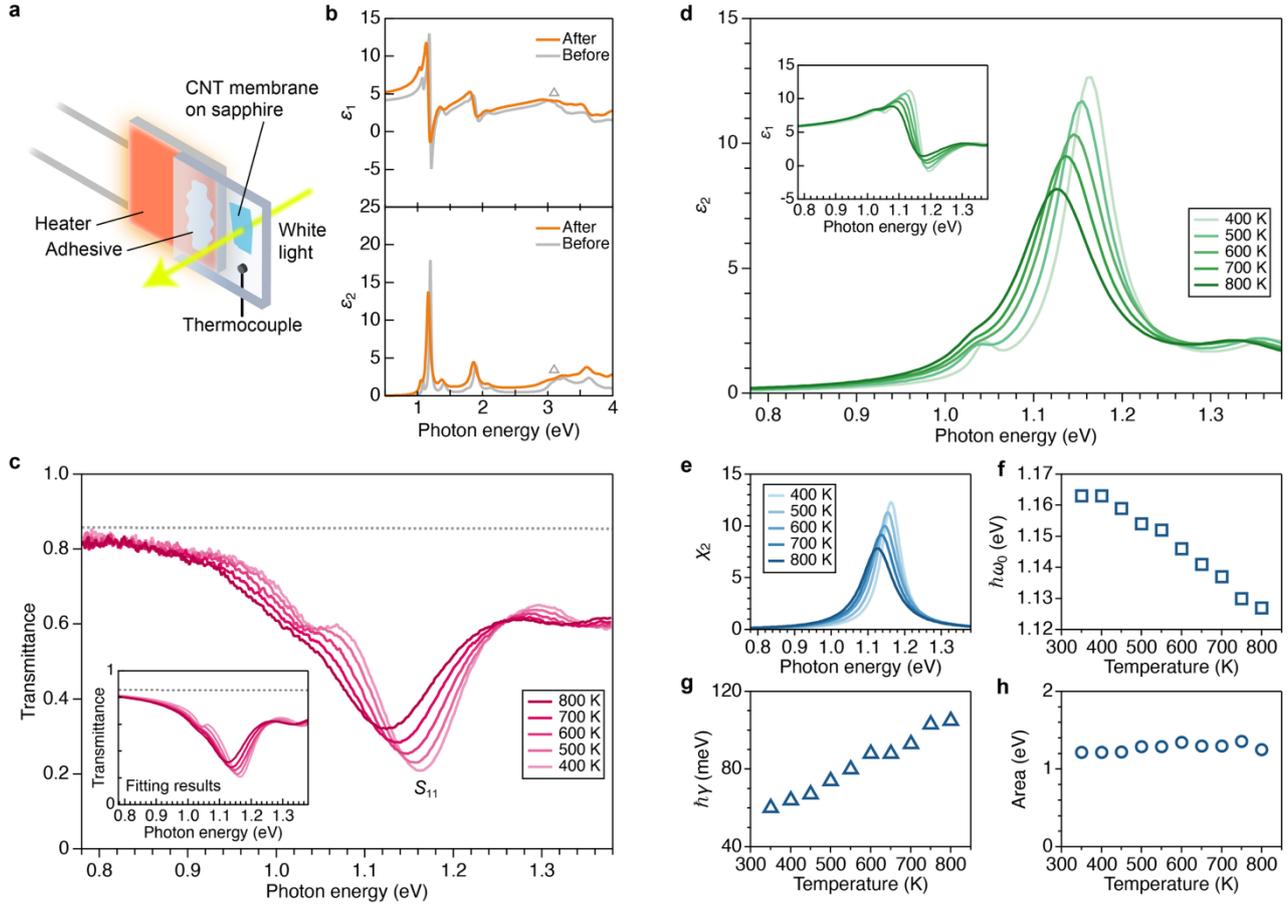

**Fig. 3 | Temperature dependence of transmittance. a** Schematic of the optical setup for high-temperature transmittance measurements of the transmittance of carbon nanotube (CNT) membrane. **b** Complex dielectric function ($\tilde{\varepsilon} = \varepsilon_1 + i\varepsilon_2$) of the (7,5) CNT membrane before (gray) and after (orange) three heating–cooling cycles. Open triangles indicate the absorption peak position of the polymers. **c** Temperature dependence of the transmittance spectra around the first subband ($S_{11}$) exciton and the sapphire substrate (gray dotted lines), along with the fitting results (inset). Temperature dependences of **d** $\varepsilon_1$ (inset) and $\varepsilon_2$ obtained from the fittings and **e** the imaginary part of the optical susceptibility ($\tilde{\chi} = \chi_1 + i\chi_2$) of the $S_{11}$ exciton of (7,5) CNTs, which can be described by a Lorentzian function $\tilde{\chi}(\omega) = f[(\omega_0^2 - \omega^2) - i\omega\gamma]^{-1}$ with $f$, $\omega_0$, and $\gamma$ being the strength, resonant optical frequency, and damping term, respectively. Temperature dependence of **f** $\hbar\omega_0$ ($\hbar$ = reduced Planck constant), **g** $\hbar\gamma$, and **h** area.

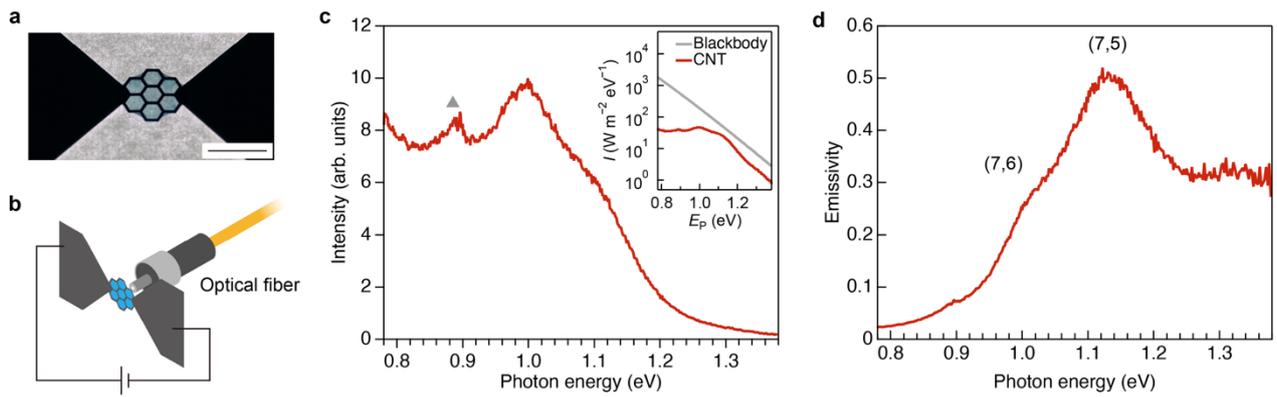

**Fig. 4 | Thermal radiation of the carbon nanotube membrane. a** Photograph of the carbon nanotube (CNT) membrane on the specially designed tungsten plate (scale bar = 5 mm). **b** Schematic of the optical setup for measuring thermal radiation. **c** Thermal radiation spectrum and **d** absolute spectral emissivity of the CNT membrane. In **c**, the gray triangle indicates the artifact from the optical fiber and the inset shows the spectra of the CNT membrane and a black body at 850 K (log–linear scale). $I$ and $E_P$ denote the intensity and photon energy, respectively.

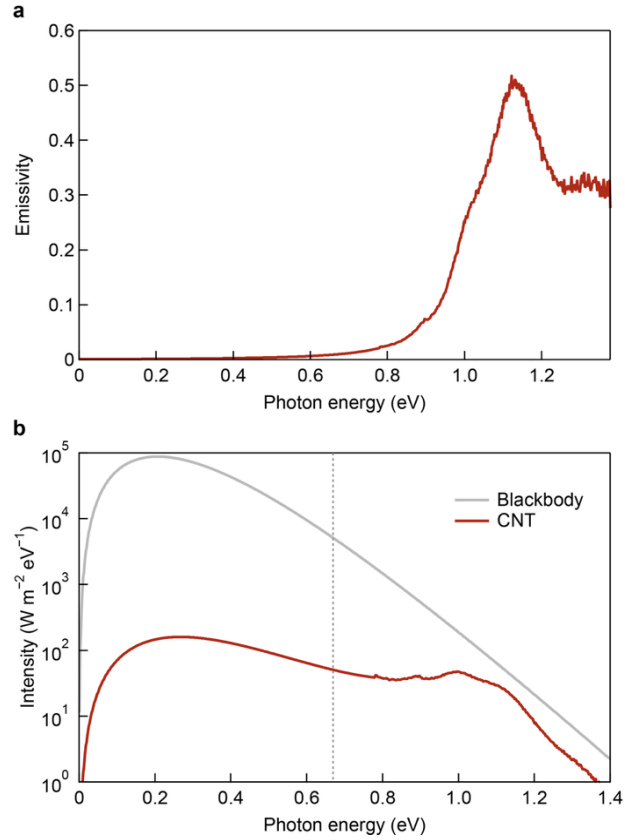

**Fig. 5 | Broadband optical spectra. a** The absolute spectral emissivity of the free-standing (7,5) carbon nanotube (CNT) membrane at 850 K. **b** Calculated thermal radiation spectra of the CNT membrane and a blackbody at 850 K (logarithmic–linear plot).

**Supplementary information**

**Exciton thermal radiation from structure-sorted carbon nanotube membranes**

Akiteru Takahashi[1†], Kaichi Teranishi[1†], Shonosuke Takaichi[1], Taishi Nishihara[1,2]*, Yuhei Miyauchi[1]*

[1]*Institute of Advanced Energy, Kyoto University, Kyoto 611-0011, Japan*

[2]*Department of Physics, Tokyo University of Science, Kagurazaka 1-3, Shinjuku, Tokyo 162-8601, Japan*

**Supplementary Note 1: Complex dielectric function spectrum**

The complex dielectric function spectrum of the single-walled carbon nanotube (SWCNT) membrane was determined with reference to a previous work[1]. First, the microscopic optical susceptibility $\tilde{\chi}$ of an SWCNT membrane of bulk density $\rho$ was modeled as follows:

$$\tilde{\chi}(\rho, \omega) = \rho \left[ \sum_i \tilde{\chi}_L^i(\omega) + \tilde{\chi}_C(\omega) + \tilde{\chi}_D(\omega) + \chi_B \right], \quad (1)$$

where $\omega$ is the optical frequency.

The first term on the right-hand side of Eq. (1), $\tilde{\chi}_L^i$, is a Lorentzian function of the $i$th peaked oscillator (where the oscillators correspond to excitons and phonon sidebands). It is calculated as follows:

$$\tilde{\chi}_L^i(\omega) = f_L^i \left[ \left( {\omega_L^i}^2 - \omega^2 \right) - i\omega \gamma_L^i \right]^{-1},$$

where $f_L^i$, $\omega_L^i$, and $\gamma_L^i$ are the strength, resonant optical frequency, and damping term of the $i$th Lorentz oscillator, respectively.

The second term $\tilde{\chi}_C$ on the right-hand side of Eq. (1), represents the nearly featureless continuum band in the photon-energy region above the first subband ($S_{11}$) exciton resonance. This term is given by a phenomenological step-like function as follows:

$$\tilde{\chi}_C(\omega) = A_C \left[ G\big((\omega - \omega_C)\gamma_C^{-1}\big) + G\big((\omega + \omega_C)\gamma_C^{-1}\big) - G((\omega - \omega_{cut})\gamma_{cut}^{-1}) - G((\omega + \omega_{cut})\gamma_{cut}^{-1}) \right],$$

where $G(x)$ is a series of $\pi/2$-phase-shifted Lorentzian functions $g(x) = (x + i\theta)^{-1}$; in particular, $G(x) = \sum_{p=0}^{100} 2ig(x, (p + 0.5)\pi)$, where $p$ is the summation index; $A_c$ is the absorption strength; and $\omega_c$ and $\gamma_C^{-1}$ are parameters specifying the center frequency and gradient at $\omega = \omega_C$, respectively. The cutoff energy and gradient are given by $\hbar\omega_{cut} = 6$ eV and $(\hbar\gamma_{cut})^{-1} = 1$ eV$^{-1}$, respectively.

The third term $\tilde{\chi}_D$ on the right-hand side of Eq. (1) represents the Drude response of free carriers. We set $\tilde{\chi}_D(\omega) = 0$ because our membrane showed no Drude response at lower photon energies (Fig. 2d). The fourth term $\chi_B$ on the right-hand side of Eq. (1) is the background susceptibility (a real constant) describing all contributions other than the abovementioned oscillators. The corresponding complex relative dielectric function ($\tilde{\varepsilon}$) is given as $\tilde{\varepsilon}(\rho, \omega) = 1 + \tilde{\chi}(\rho, \omega)$.

The reflection and transmission spectra of the SWCNT membrane with $\tilde{\varepsilon}(\rho, \omega)$ were calculated using the optical transfer matrix technique[2]. For the on-sapphire SWCNT membrane, the back surface reflection must also be considered[1]. Here, $\rho$ was set to 1 g cm$^{-3}$ (the typical density of SWCNT membranes[1]) and $\tilde{\varepsilon}(\rho, \omega)$ was obtained by determining the above oscillator parameters that reproduce all reflectance and transmittance spectra of the SWCNT membrane (Fig. 2d).

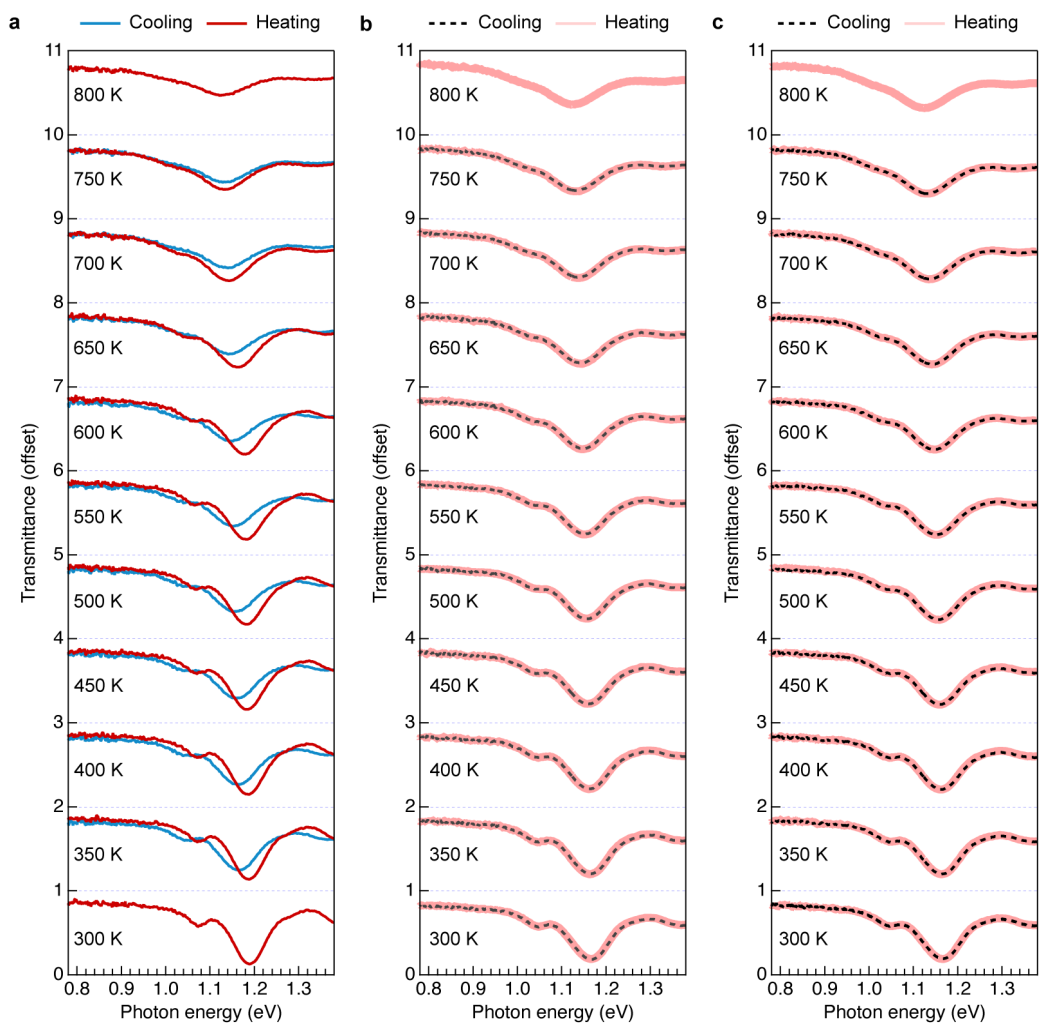

**Supplementary Fig. 1 | Temperature dependences of the transmittance spectra.** The transmittance spectra of the (7,5) carbon nanotube membrane are shown during the **a** first, **b** second, and **c** third heating (from 300 K to 800 K) and cooling (from 800 to 300 K) processes. The spectra are offset for visualization purposes (the dotted line in each spectrum indicates the zero line).

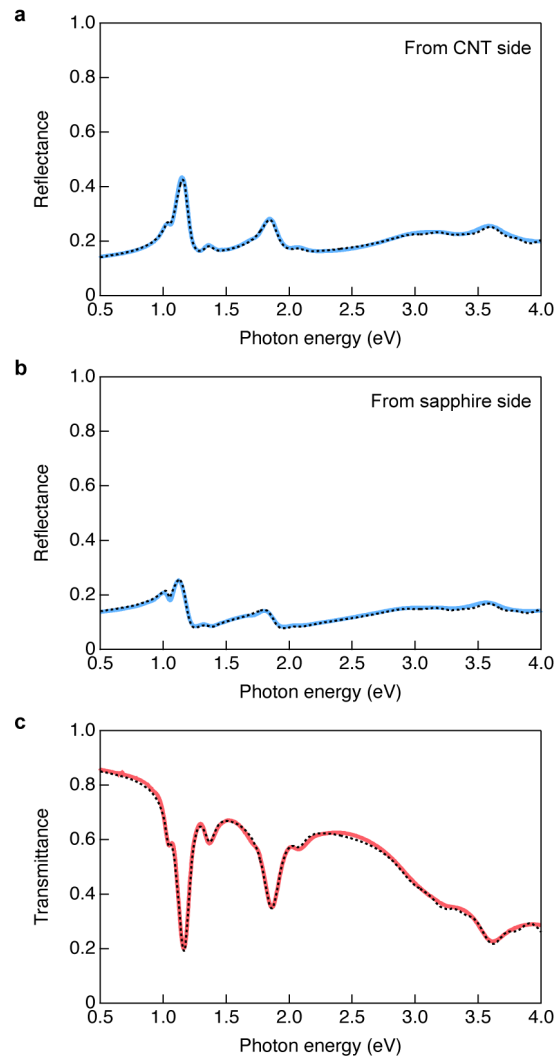

**Supplementary Fig. 2 | Optical spectra of the carbon nanotubes after the heating–cooling cycles.**
**a, b** Reflectance and **c** transmittance spectra of the (7,5) carbon nanotube (CNT) membrane after three heating–cooling cycles. The reflectance spectra were obtained in two configurations: **a** incident light entering from either side of the CNT membrane and **b** incident light entering the sapphire substrate. The dotted black curves are the fitting results.

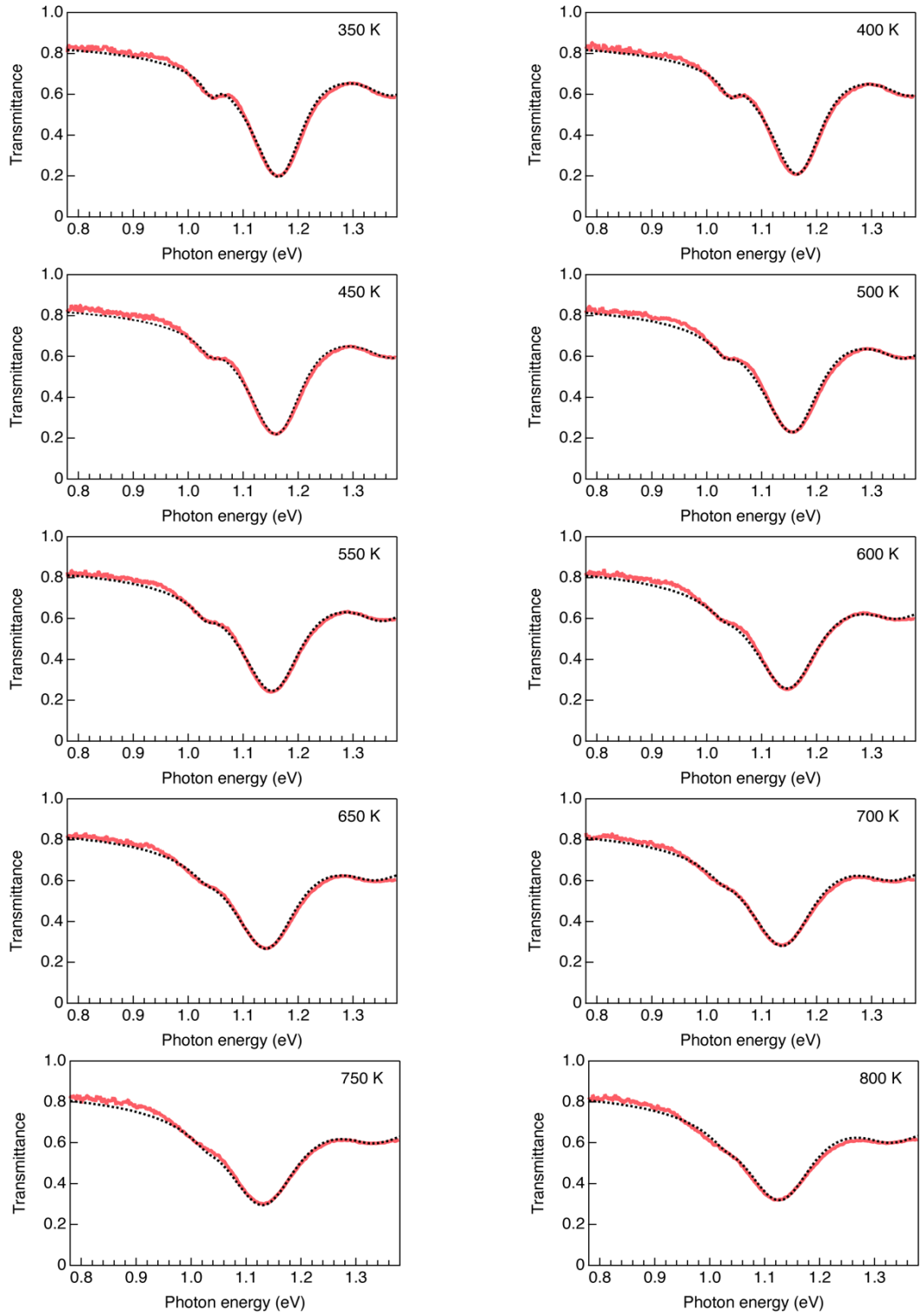

**Supplementary Fig. 3 | Transmittance spectra and their fitting results.** The transmittance spectra (red curves) of the (7,5) CNT membrane are shown during the third cooling process obtained at different heating temperatures (350 K–800 K). The black dotted curves are the fitting results.

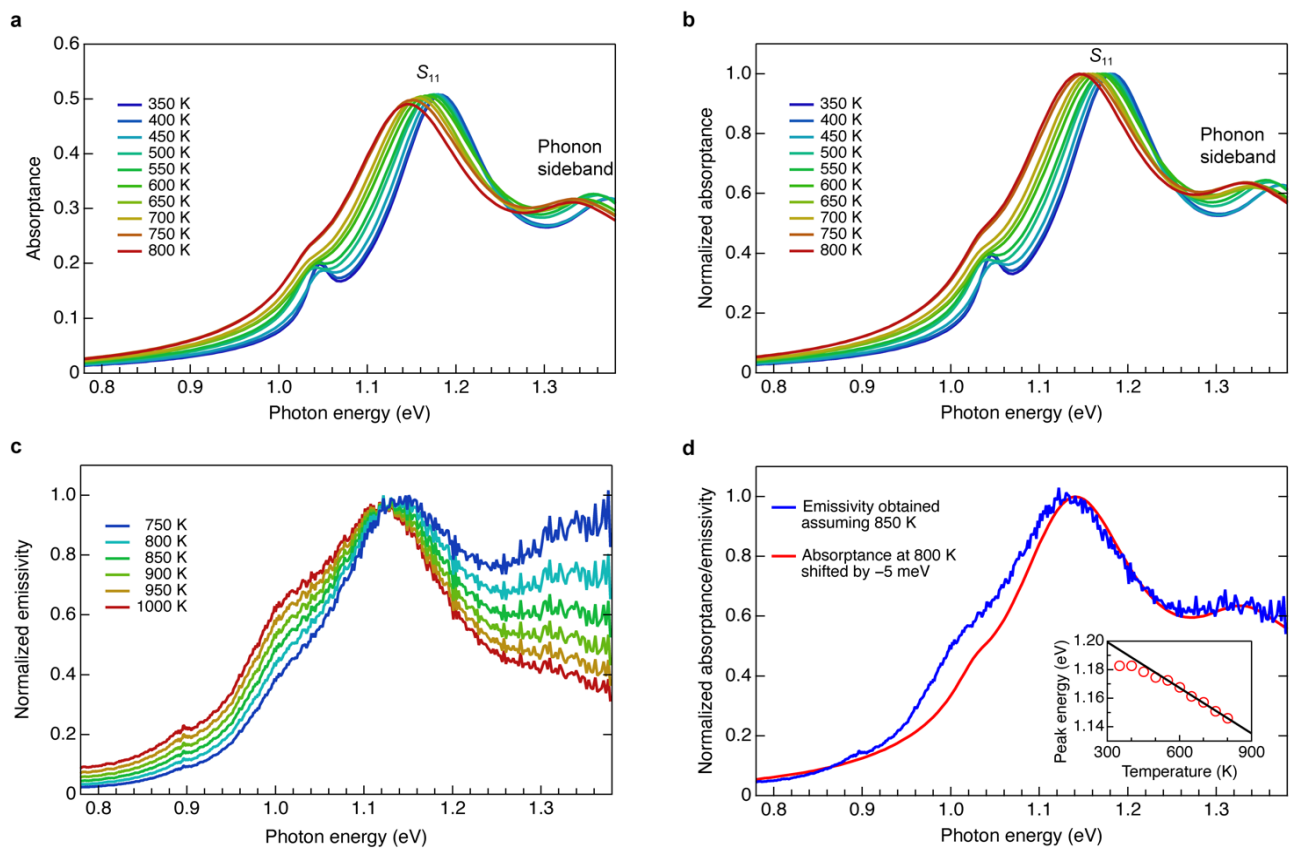

**Supplementary Fig. 4 | Temperature estimation from the absorptance and emissivity spectra. a** Absorptance spectra of the free-standing SWCNT membrane at different temperatures calculated from the complex dielectric function of the on-sapphire membrane (Fig. 2). **b** Absorbance spectra of the free-standing SWCNT membrane normalized at the absorbance intensity of the first subband ($S_{11}$) exciton. **c** Candidate emissivity spectra of the free-standing SWCNT membrane emitting thermal radiation (Fig. 4c) assumed at various temperatures (all intensities are normalized at the $S_{11}$ exciton resonance 1.1–1.2 eV). **d** Comparison of the emissivity spectra obtained at an assumed temperature of 850 K and the absorbance at 800 K (left-shifted by 5 meV to correct for the 50 K temperature difference). The inset plots the temperature dependence of the absorbance peak of the $S_{11}$ exciton (from panel **a**) and its linear fitting result (black line).